\documentclass[12pt, a4paper]{elsarticle}
\usepackage{amssymb}
\usepackage{hyperref}
\usepackage{multirow}
\usepackage{caption}
\usepackage{subcaption}
\usepackage{paralist}
\usepackage[disable]{todonotes}

\usepackage{listings}
\usepackage{xcolor}

\definecolor{comments}{rgb}{0.48,0.71,0.92}
\definecolor{codegray}{rgb}{0.5,0.5,0.5}
\definecolor{codestring}{rgb}{0.19,0.28,0.52}
\definecolor{backcolour}{rgb}{0.98,0.98,0.97}

\lstdefinestyle{mystyle}{
    basicstyle=\scriptsize\ttfamily,
    backgroundcolor=\color{backcolour},   
    commentstyle=\color{comments},
    numberstyle=\tiny\color{codegray},
    stringstyle=\color{codestring},
    breakatwhitespace=false,         
    breaklines=true,                 
    captionpos=b,                    
    keepspaces=true,                 
    numbers=left,                    
    numbersep=5pt,                  
    showspaces=false,                
    showstringspaces=false,
    showtabs=false,                  
    tabsize=1
}

\usepackage{float}

\restylefloat{table}

\journal{arxiv}

\begin{document}

\begin{frontmatter}



\title{SofaMyRoom: a fast and multiplatform ``shoebox'' room simulator for binaural room impulse response dataset generation.}

\author[1]{Roberto Barumerli\corref{cor1}}
\author[2]{Daniele Bianchi}
\author[3]{Michele Geronazzo}
\author[2]{Federico Avanzini}

\address[1]{Acoustic Research Institute, Austrian Academy of Sciences, Vienna, Austria}
\address[2]{Dept. of Computer Science Dept., University of Milan, Milan, 20135, Italy}
\address[3]{Dyson School of Design Engineering, Imperial College London, London, SW7 2AZ, United Kingdom}
\cortext[cor1]{Corresponding author. 
E-mail address: roberto.barumerli@oeaw.ac.at}
\begin{abstract}
%\paper{ Please, keep in mind the core philosophy of the journal: the actual publication is the code/software itself. The manuscript you submit must be regarded as a brief accompanying note.}
This paper introduces a shoebox room simulator able to systematically generate synthetic datasets of binaural room impulse responses (BRIRs) given an arbitrary set of head-related transfer functions (HRTFs). The evaluation of machine hearing algorithms frequently requires BRIR datasets in order to simulate the acoustics of any environment. However, currently available solutions typically consider only HRTFs measured on dummy heads, which poorly characterize the high variability in spatial sound perception. Our solution allows to integrate a room impulse response (RIR) simulator with different HRTF sets represented in Spatially Oriented Format for Acoustics (SOFA). The source code and the compiled binaries for different operating systems allow to both advanced and non-expert users to benefit from our toolbox.

\end{abstract}


\end{frontmatter}

\section*{Required Metadata}

\section*{Current code version}

\begin{table}[H]
\hspace{-2cm}
\begin{tabular}{|l|p{6.5cm}|p{8.5cm}|}
\hline
\textbf{Nr.} & \textbf{Code metadata description} & \textbf{Please fill in this column} \\
\hline
C1 & Current code version & v1.0 \\
\hline
C2 & Permanent link to code/repository used for this code version &  \url{https://github.com/spatialaudiotools/sofamyroom} \\
\hline
C3 & Code Ocean compute capsule & none \\
\hline
C4 & Legal Code License & EUPL v1.2 \\
\hline
C5 & Code versioning system used & git \\
\hline
C6 & Software code languages, tools, and services used & C, MATLAB (optional), python3 (docs), Visual Studio C IDE (optional). \\
\hline
C7 & Compilation requirements, operating environments \& dependencies & gcc, cmake\\
\hline
C8 & Documentation &  \url{https://spatialaudiotools.github.io/sofamyroom/} \\
\hline
C9 & Support email for questions & roberto.barumerli@oeaw.ac.at \\
\hline
\end{tabular}
\caption{Code metadata}
 
\end{table}

%\linenumbers

%\paper{
%The permanent link to code/repository or the zip archive should include the following requirements: 
%
%README.txt and LICENSE.txt.
%
%Source code in a src/ directory, not the root of the repository.
%
%Tag corresponding with the version of the software that is reviewed.
%
%Documentation in the repository in a docs/ directory, and/or READMEs, as appropriate.}

%\rb{checklist https://www.elsevier.com/journals/softwarex/2352-7110/guide-for-authors}
%====================================
\section{Motivation and significance}
\label{sect:motivation}
%\paper{Introduce the scientific background and the motivation for developing the software.}
%\paper{Introduce related work in literature (cite or list algorithms used, other software etc.).}
%\paper{Explain why the software is important, and describe the exact (scientific) problem(s) it solves.}
%\paper{Provide a description of the experimental setting (how does the user use the software?).}
The process required to record binaural room impulse responses (BRIRs) with peculiar acoustic proprieties is expensive and time consuming~\cite{kuttruff2016room}. These responses are the combination of the temporal dynamic of an echoic environment, named room impulse response (RIR), with the individual spatial filtering effects related to the binaural acoustic transformation of a human subject, typically summarized by head-related impulse response (HRIR), or equivalently by their Fourier-transformed head-related transfer functions (HRTFs)~\cite{blauert2013technology}. Given the difficulties involved in the acquisition of BRIRs, the introduction of room acoustic simulators~\cite{allen_image_1979} allowed to generate synthetic responses and, consequently, new approaches emerged in psycho-acoustics and machine hearing~\cite{lyon2017human} research. 
\todo[inline, color=yellow]{DO NOT OVERSTATE: we are not able to handle real acoustic complexity\\
In recent years, psycho-acoustic studies demonstrated the perceptual validity of such synthetic reverberation~\cite{weinzierl_measuring_2018}, introducing the possibility to test subjects in complex listening situations otherwise difficult to fabricate in a real setup. Such virtual sound environments can be adopted in several context, such as: auralization schemes~\cite{zahorik_perceptually_2009}, hearing aid fitting or hearing loss simulations ~\cite{mueller_localization_2012}, cochlear implants processing methods~\cite{lopez-poveda_lateralization_2019}, and so on.}

Concerning psycho-acoustic studies, designing synthetic BRIRs requires to consider specific stimulus conditions, since human listeners successfully tackle adverse acoustic conditions by solving the so-called \textit{cocktail party} problem~\cite{haykin2005cocktail}: consequently, these experimental designs evaluate limited amount of cases to test precise components of the human behaviour.
On the other hand, inspired by the human hearing capabilities, several computational approaches known as ``machine hearing'' have been proposed ~\cite{lyon2017human}. This approaches incorporate audio signal processing stages that simulate the functioning of the human auditory system in combination with machine learning algorithms to address problems like: speech de-reverberation, sound source separation, and sound localization in acoustically reverberant environments. The resulting software frameworks require large amounts of annotated data for training and: while for well-established topics, such as speech recognition, a plethora of speech corpora is publicly available~\cite{gannot2017consolidated}, within the field of computational auditory scene analysis (CASA)~\cite{brown1994computational} the availability of public datasets is limited to a few specific cases. 

Currently available datasets have two main limitations: (i) a small amount of echoic conditions are available; (ii) binaural recordings are obtained using mannequins (e.g., the Neumann NU100 dummy-head or the GRAS Kemar head-torso simulator~\cite{gardner_hrtf_1995}), which do not account for the variability of human HRTFs due to individual anthropometric differences.
Such restrictions can be overcome by relying on numerical BRIR simulation frameworks,
%. While this approach can be limited to describe the variabilities of a real scenario, it permits 
which permit to automatically generate large amounts of annotated data over disparate conditions. Commercial solutions exist, such as Odeon\footnote{\url{https://odeon.dk}} or EASE.\footnote{\url{http://ease.afmg.eu}} While these solutions are able to render complex environments with high accuracy, they do not always guarantee the flexibility needed for research purposes (i.e. limited support to HRTF datasets). Also several academic projects are available but no one of them allow to render BRIRs with a specific HRTF dataset in a straighforward and iterateable manner.
%. These last simulators were not included for the presence of peculiar incompatibilities such as: the scarse of available documentation within the EVERTims project, RAVEN's code is not released, 
%the authors extended the previous toolbox adding the feature to support microphone arrays but the code is not available and its use is limited as a MATLAB library. 

In order to overcome the above limitations, this work proposes \texttt{SofaMyRoom}, a toolbox for BRIR simulation developed around three key components: (i) efficient and realistic acoustic simulations, (ii) high flexibility to describe listener acoustics through individual HRTFs, and (iii) automatic annotation of simulated responses. 
%\the\ttextwidth
\begin{figure*}[t!]
     \centering
     \begin{subfigure}[l]{0.48\textwidth}
     \centering
        \includegraphics[width=\textwidth]{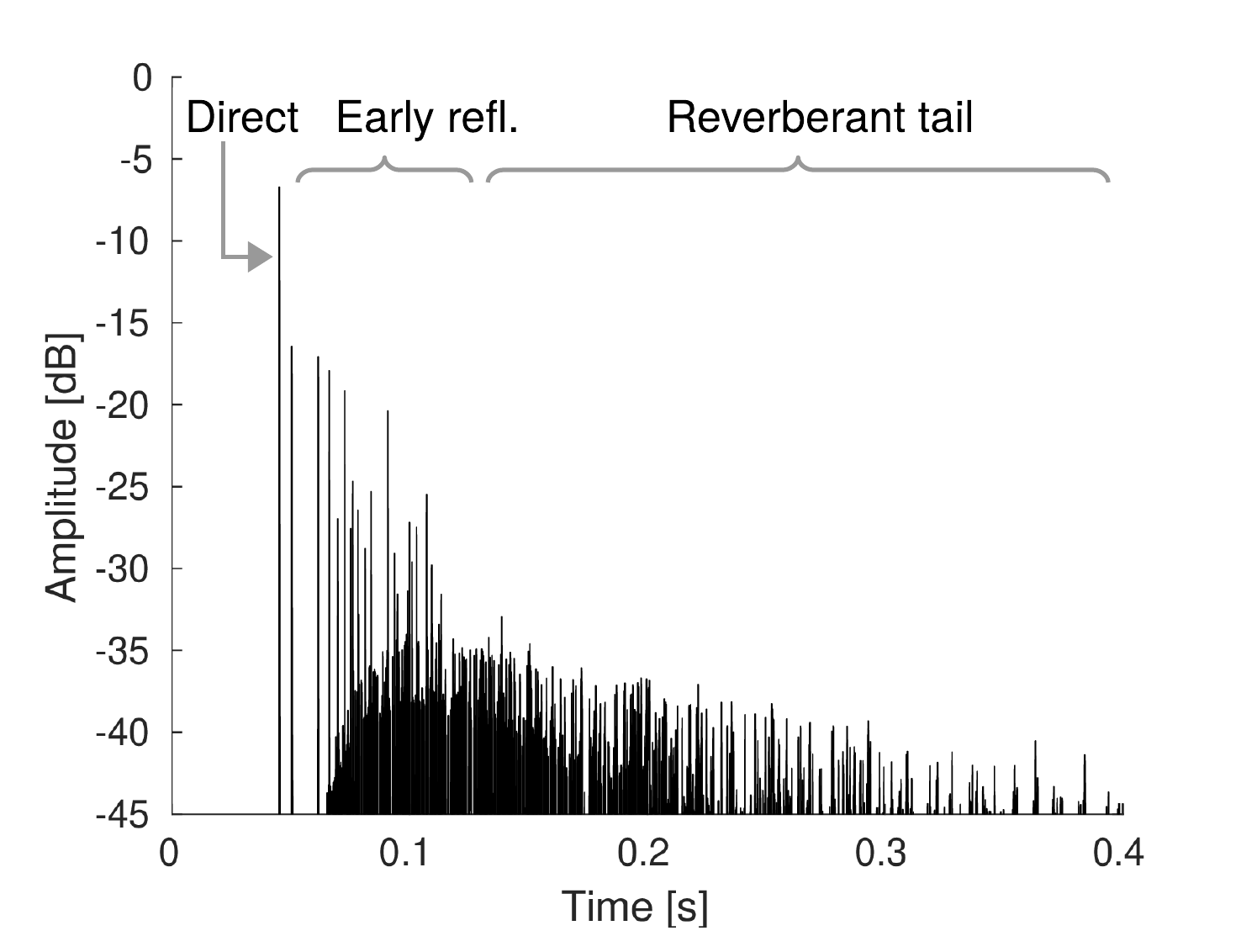}
        \caption{A RIR and its main parts: direct component, early reflections and reverberant tail.}
        \label{fig:rir_comp}
    \end{subfigure}
    \hfill
    \begin{subfigure}[r]{0.48\textwidth}
    \centering
    \includegraphics[width=\textwidth]{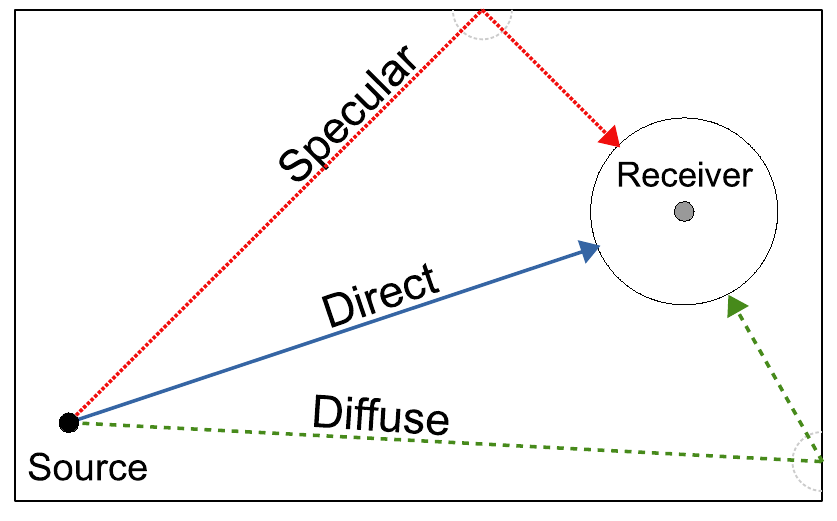}
    \caption{Spatial evolution. Direct (blue), and specular (red) paths are generated through the image source method, while diffused rays (green) are generated through the diffuse rain algorithm. The circle around the receiver identifies its detection sphere.}
    \label{fig:methods}
    \end{subfigure}
    \caption{Schematic representation of the spatio-temporal evolution of a simulated RIR.}
\end{figure*}

The \texttt{SofaMyRoom} acoustic simulator is based on the original \texttt{roomsim} project~\cite{schimmel_fast_2009}. %Although shoe-box shaped rooms have limited generality, ``roomsim'' provides an efficient solution for the use cases outlined at the beginning of this section. 
Such toolbox combines the image-source method~\cite{allen_image_1979} and the diffuse rain ray-tracing algorithm~\cite{heinz_binaural_1993} to account for the three main parts of a RIR~\cite{kuttruff2016room} (see Fig.~\ref{fig:rir_comp}): %\cite{Tsilfidis2013}
(i) the direct component, or the line of sight propagation; (ii) the early (specular) reflections, which the human auditory system is able to leverage to extract information (i.e. improving speech intelligibility or distance perception); (iii) the reverberant tail, which aggregates all the late diffused and reflected wave-paths and typically contains relevant information about the size and materials of the room. 

HRTF individualization makes use of the SOFA format, originally proposed by Majdak~\cite{majdak_spatially_2013} and standardized by the Audio Engineering Society (AES) as AES69-2015 in 2015. Prior to SOFA, HRTF data were stored according to custom file formats, thus making the data exchange difficult~\cite{geronazzo_standardized_2013}. By employing SOFA, \texttt{SofaMyRoom} enables to load into the simulation a vast amount of HRTF datasets already stored in this format.\footnote{\url{https://www.sofaconventions.org/data/database}} 

Finally, we added the possibility to automate BRIR generation and annotation through the Virtual Acoustic Space Traveler (VAST) project~\cite{gaultier2017vast}. This provides the user with a simple and effective tool to systematically generate BRIR datasets accompanied by a metadata structure which describes each of the rendered samples.\footnote{For the VAST structure check the documentation or visit \url{http://thevastproject.inria.fr/}}
As a result, with a few lines of code \texttt{SofaMyRoom} can synthesise annotated datasets of BRIRs through the parametrization of geometric and acoustic settings. Finally, the standardized metadata structure simplifies the data integration into the machine learning pipeline~\cite{dong_data_2018}. This last feature is especially desirable, since also open datasets could be organized with the VAST structure, which can support faster hypothesis testing and rapid prototyping. 

%====================================
\section{Software description}
\label{sect:desc}
%\paper{Describe the software in as much as is necessary to establish a vocabulary needed to explain its impact. }
%

\texttt{SofaMyRoom} is written in standard C, where all the libraries are statically linked\footnote{Except for FFTW3 on Windows which has been linked dynamically.} providing a ready to use multi-platform executable. In order to guarantee full portability, we choose CMAKE as building system which can compile the executable along with the documentation and  expose the unit tests. \texttt{SofaMyRoom} can be executed on different platforms as a native software by passing simulation parameters as standard input or a text file. The user is then able to render multiple BRIRs by defining one single room and multiple receivers and sources, and to save them into separate 2-channel WAVE files.
The adopted license, EUPL v1.2, and the integration of CMAKE, allow advanced users to customize and extend \texttt{SofaMyRoom}'s code.
%with the adoption of the preferred editor tool (i.e. Microsoft Visual Studio Code or Community). 

\subsection{Architecture}
\label{sect:architecture}
\begin{figure*}[t!]
    \centering
    \begin{subfigure}[t]{0.5\textwidth}
        \centering
        \includegraphics[width=\textwidth]{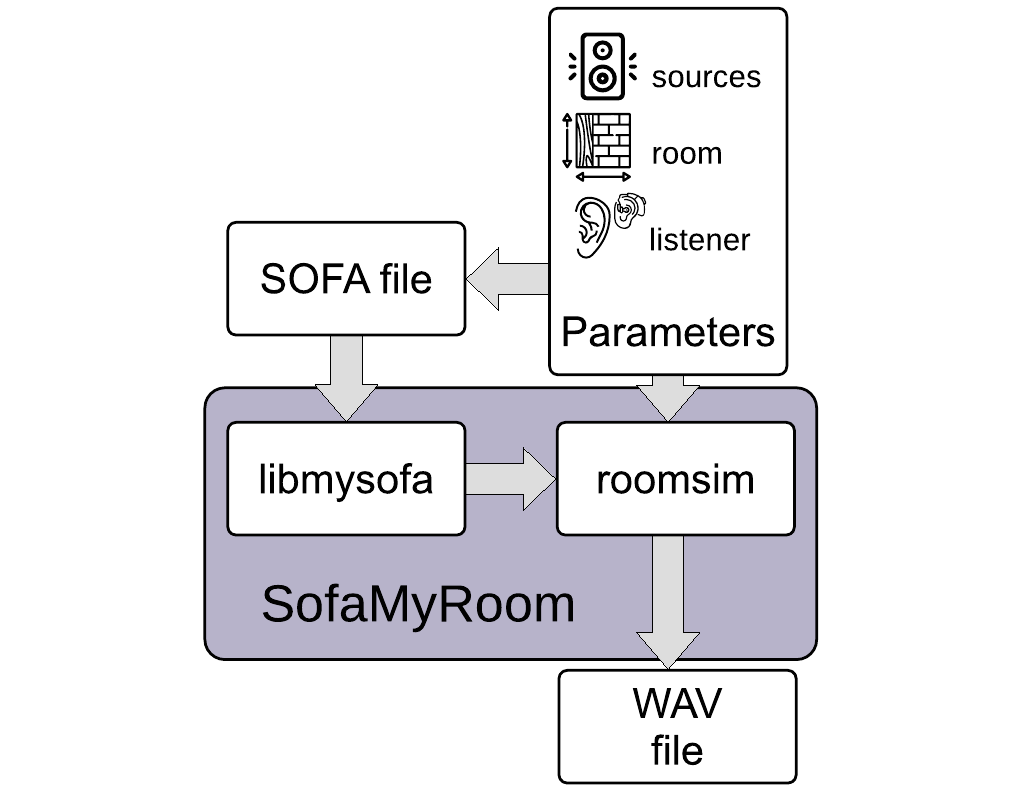}
        \caption{Single BRIR file generation.}
        \label{fig:wavgen}
    \end{subfigure}%
    \begin{subfigure}[t]{0.5\textwidth}
        \centering
        \includegraphics[width=\textwidth]{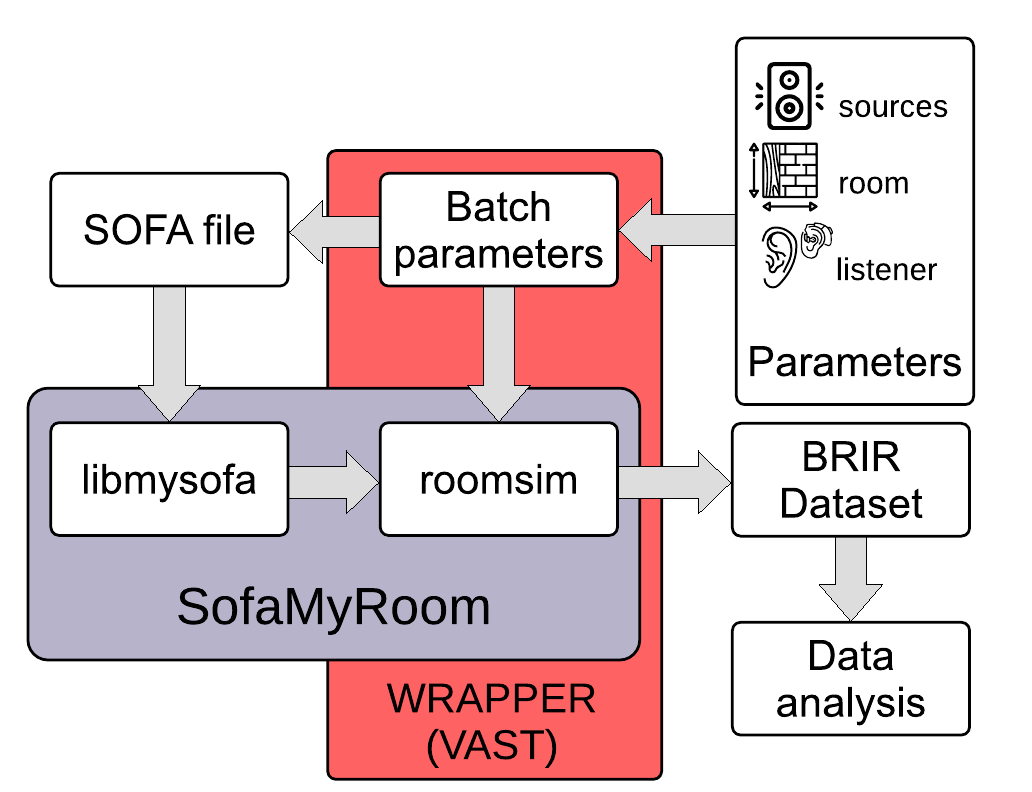}
        \caption{Workflow example for the BRIR dataset generation and analysis using the \texttt{SofaMyRoom} wrapper. }
        \label{fig:setgen}
    \end{subfigure}
\caption{Block diagram of \texttt{SofaMyRoom}'s structure and  workflows.} 
\label{fig:sofamyroomschema}
\end{figure*}
%\paper{Give a short overview of the overall software architecture; provide a pictorial component overview or similar (if possible). If necessary provide implementation details.}

Figure \ref{fig:sofamyroomschema} provides a block-diagram of the software architecture. It can be noted that \texttt{SofaMyroom} is built by integrating in the same package a refined and updated version of the acoustic simulator of the discontinued \texttt{roomsim} project released by Schimmel et al.~\cite{schimmel_fast_2009}, and \texttt{libmysofa}, a Standard C library to load and handle SOFA files~\cite{hoene_mysofa_2017}. Moreover, in order to provide the capability to generate an annotated BRIR dataset we integrated a refactored and documented version of the VAST project~\cite{gaultier2017vast}. Since VAST has been developed as a MATLAB toolbox, we released \texttt{SofaMyRoom} also as a MATLAB function via MEX compilation.

As showed in Figure \ref{fig:sofamyroomschema}, two possible workflows are considered: the generation of either a single BRIR or an entire BRIR dataset.
In the first case (Fig.~\ref{fig:wavgen}), the user may need to evaluate a specific acoustic configuration of room, source and listener as in psycho-acoustic experiments. In the second case (Fig.~\ref{fig:setgen}), which is more typical for the generation of training data in a machine hearing scenario, several simulations need to be defined and generated in batch.

% \subsection{Simulation methods}

\subsection{Functionalities}
\label{sect:functionalities}
%\paper{Present the major functionalities of the software.}

For each simulation, the user can specify the acoustic parameters of an empty ``shoebox'' reverberant environment. The main elements are: the room size, the frequency-dependent absorption and scattering coefficients for each of the six rectangular surfaces, and the source and receiver positions. Additional parameters can control the simulator behaviour and output such as: sampling frequency, room temperature and humidity, the impulse response maximum duration, the frequency bands to be accounted.\footnote{The complete set of parameters is specified in the software documentation}

Sound sources can be specified through their directivity pattern (i.e., omnidirectional or cardiod) while receivers can be characterized by a set of HRTFs. \texttt{SofaMyRoom} allows to handle HRTFs in SOFA format providing high flexibility in defining the receiver's geometry and behaviour, with potentially unlimited configurations. With the integration of the \texttt{libmysofa} package, our toolbox can not only read data stored in a file according to the standard, but also manipulate them. HRTF data can be normalized upon opening, or resampled according to the project sampling frequency. Moreover, the HRTF corresponding to an arbitrary direction can be computed through its closest available datapoint, or through bilinear interpolation. 

The user is also able to select which algorithm should be used for the acoustic simulation by enabling or disabling the image source method or diffuse rain algorithm. 
Simulation results are eventually saved on disk according to the user preferences: as a MATLAB array or a WAVE file. % with the user's specified name. 

Finally, the integration with the VAST toolbox~\cite{kataria_hearing_2017} allows the user to easily generate arbitrarily large datasets of BRIRs. The provided MATLAB scripts allow to: (i) initialize an empty VAST structure, (ii) define the room acoustic conditions and (iii) automatically populate the dataset with meta-data while calling \texttt{SofaMyRoom} to generate and store BRIRs. While the current version of the script that populates the VAST structure only allows to iterate over user-defined receivers' positions inside the room, the published code can be easily extend to integrate other conditions (i.e. diverse HRTF datasets).

\subsection{First-time setup}
\label{sect:sample}
The minimal setup for \texttt{SofaMyRoom} requires three files: % into the same folder: 
\begin{itemize}
    \item the \texttt{SofaMyRoom} executable;
    \item 1 text file with simulation parameters (e.g.,  \texttt{setup.txt});
    \item a HRTF set wrapped into a SOFA file (e.g., \texttt{subject\_003.sofa}\footnote{Available at \url{https://www.sofaconventions.org/data/database/cipic/subject\_003.sofa}}).
\end{itemize}
An example of parameters' file is reported in Listing~\ref{paramfile} (the diffuse acoustic field is omitted for brevity).

\begin{lstlisting}[language=Octave, label={paramfile}, caption={Example of a parameter file.} ]
% room settings
room.dimension = [10 7 4]; % [m]
room.humidity = 0.42; % relative humidity 
room.temperature = 20; % [deg C]

% diffuse reflections simulation options
options.simulatediffuse = false; 

% specular reflections simulation options
options.simulatespecular = true; 
options.reflectionorder = [ 10 10 10 ]; 

% surface coefficients
room.surface.frequency = [125 250 500 1000 2000 4000]; % [Hz]
room.surface.absorption = [0.1000 0.0500 0.0600 0.0700 0.1000 0.1000;
				0.1400 0.3500 0.5300 0.7500 0.7000 0.6000;
				0.1000 0.0500 0.0600 0.0700 0.1000 0.1000;
				0.1000 0.0500 0.0600 0.0700 0.1000 0.1000;
				0.0100 0.0200 0.0600 0.1500 0.2500 0.4500;
				0.2400 0.1900 0.1400 0.0800 0.1300 0.1000 ];

% simulation options
options.fs = 44100; % [Hz] 
options.responseduration = 1.25; % [s]
options.bandsperoctave = 1; %  [bands/octave]
options.referencefrequency = 125; %  [Hz]
options.airabsorption = true;
options.distanceattenuation = true; 
options.subsampleaccuracy = false;
options.highpasscutoff = 0; 
options.verbose = true; 

% output options
options.outputname = 'output'; 	

% source definitions
source(1).location = [ 8 2.5 1.6 ]; % [m]
source(1).orientation = [ 180 0 0 ]; % [deg]
source(1).description = 'subcardioid'; 

% receiver definitions
receiver(1).location = [ 3 5 1.2 ]; % [m]
receiver(1).orientation = [ 0 0 0 ]; % [deg]
receiver(1).description = 'SOFA cipic_subject_003.sofa'; 
\end{lstlisting}

Simulations can then be run with the following command (from a Bash interpreter):
\begin{lstlisting}[language=Bash]
~$ ./sofamyroom setup.txt
\end{lstlisting}

Assuming the correctness of all the parameters, %otherwise an error will be reported on screen, 
BRIRs are stored into WAVE files ready to be used for further processing.
%While this generation process outputs few BRIRs at a time, the synthetic samples can be used to run a psycho-acoustic experiment to evaluate the human perception over diverse acoustic conditions (i.e.~\cite{mueller_localization_2012}).

%===================================
\section{Illustrative Examples}
\label{sect:example}
%\paper{Provide at least one illustrative example to demonstrate the major functions. Optional: you may include one explanatory video that will appear next to your article, in the right hand side panel. (Please upload any video as a single supplementary file with your article. Only one MP4 formatted, with 50MB maximum size, video is possible per article. Recommended video dimensions are 640 x 480 at a maximum of 30 frames/second. Prior to submission please test and validate your .mp4 file at $ http://elsevier-apps.sciverse.com/GadgetVideoPodcastPlayerWeb/verification$. This tool will display your video exactly in the same way as it will appear on ScienceDirect.).}
% \rb{l'obiettivo di questa sezione è riportare cosa SofaMyRoom può fare e non cosa fa il modello}
Here, we propose a working example that uses a computational auditory model for sound localization~\cite{reijniers_ideal-observer_2014}, in its current implementation~\cite{barumerli_model_2020}, to demonstrate how \texttt{SofaMyRoom} can be a tool of interest for both hearing and machine audition research. The considered model mimics the performance of the human auditory system in the sound localization task, by estimating both the horizontal and vertical polar coordinates of a sound source given a binaural stimulus. The implementation is available in the Auditory Modelling Toolbox (AMToolbox),\footnote{\url{http://amtoolbox.sourceforge.net/}} a MATLAB/Octave toolbox that provides models for many processing stages of the auditory system, including outer- and middle-ear acoustics, cochlear filters, inner-hair cell models, binaural processing, and so on~\cite{sondergaard_auditory_2013}.  

Here, \texttt{SofaMyRoom} synthesizes different sets of BRIRs in order to evaluate the localization performance of the auditory model over different echoic conditions %. Since the model implementation is released within the MATLAB framework, the pipeline of the simulations is straightforward
generated by the integration of the VAST toolbox. The code snippet in Listing~\ref{vast_snippet} shows an example of dataset generation. 

\begin{lstlisting}[label={vast_snippet},caption={Illustrative example: dataset generation with VAST.}, language=Octave]
% common parameters
Fs = 16e3; [Hz]
RoomSize = [7.1 5.1 3]; % [m]
ReceiverPos = [0.211 0.294 0.5].*RoomSize; % [m]

% acoustic parameters
CeilingAbsorb = [0.02   0.06    0.14    0.37    0.60    0.65    0.65];
FloorAbsorb =   [0.55   0.86    0.83    0.87    0.90    0.87    0.87];
WallsAbsorb = CeilingAbsorb;
Diffuse = [0.1    0.1     0.1     0.1     0.1     0.1     0.1];

SofaPath = 'cipic_subject_003.sofa';
RoomName = 'room_sofa';
VAST = VASTGeneration(Fs, RoomSize, RoomName, CeilingAbsorb, FloorAbsorb, WallsAbsorb, ReceiverPos, Diffuse, SofaPath);
\end{lstlisting}

\begin{figure}[t]
	\centering
    \includegraphics[width=0.5\textwidth]{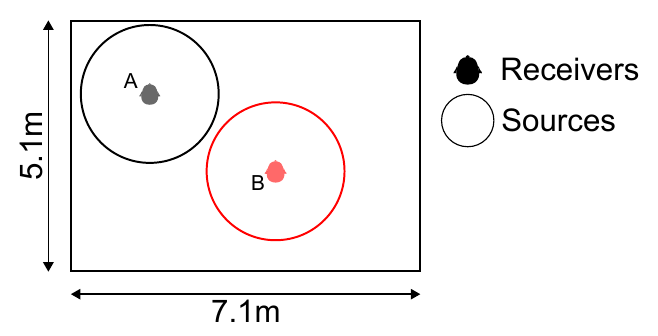}
	\caption{Top view of the simulated room. The sources are uniformly distributed along the unitary sphere with the receiver in the middle.}
	\label{fig:room}
\end{figure}

%\subsection{Simulations} 
The simulations replicate part of the experiments reported in a previous work by the authors~\cite{barumerli2019auditory}. Figure \ref{fig:room} shows a schematic representation of the configurations. Two different rooms were rendered. While the size was constant ($5.1\times7.1\times3$ ~m$^3$), two sets of absorption and scattering coefficients were adopted leading to two different Reverberation Times ($RT_{60}$): $0.37$~s, $1.88$~s. Two different positions (x,y,z) were considered for the receiver: 
\begin{compactitem}
   \item $A=(1.5, 1.5, 1.75)$ m
   \item $B=(4.05, 3.05, 1.75)$ m
\end{compactitem}
Moreover, we adopted three different HRTF datasets:\footnote{SOFA files are available at \url{www.sofaconventions.org}} 
\begin{compactitem}
\item the KEMAR mannequin from MIT~\cite{gardner_hrtf_1995};
\item subject 003 from the CIPIC database~\cite{algazi_cipic_2001};
\item subject CI1 from the ARI-BTE database~\cite{majdak_two-dimensional_2011}.
\end{compactitem}
The sound sources were broadband noise, and every direction available in the HRTF dataset was tested. The remaining parameters remained unaltered. 

The resulting set of \texttt{SofaMyRoom} simulations was straightforwardly fed to the auditory model. Although discussing the localization results achieved by the model is out of the scope of this paper, we nonetheless report them here in order to show the diversity of variables that can be investigated thanks to the flexibility of \texttt{SofaMyRoom}.

The model predictions were evaluated by means of five different perceptual metrics which quantify the precision and the localization accuracy along the horizontal and vertical coordinates~\cite{middlebrooks_virtual_1999}: lateral bias, lateral RMS error, elevation bias, local RMS polar error, and quadrant-error rate. The computation of these metrics allows to compare the model estimations with localization experiments performed with real subjects.
\begin{table*}[t]
\small
\caption{Averaged metrics comparison for two rooms, two receiver's positions, three HRTF datasets.}
\begin{center}
\begin{tabular}{|l|l|c|c|c|c|c|c|}
  \multicolumn{8}{c}{\textbf{First room} -   $RT60=0.37s$} \\
\hline
\multicolumn{2}{|r|}{HRTF dataset} & \multicolumn{2}{c|}{KEMAR MIT} & \multicolumn{2}{c|}{CIPIC SBJ3} & \multicolumn{2}{c|}{ARI-BTE CI1} \\ \hline
\multicolumn{2}{|r|}{Position} & A & B & A & B & A & B  \\ \hline \hline
\parbox[t]{2mm}{\multirow{5}{*}{\rotatebox[origin=c]{90}{Metrics}}} &
Lateral Bias $[^\circ]$             & -0.54 & 0.08 & -0.58 & 0.36 & -1.25 & 0.21 \\ \cline{2-8}
& Lateral RMS error $[^\circ]$      & 6.03 & 5.92 & 6.89 & 6.62 & 6.06 & 5.89 \\ \cline{2-8}
& Elevation bias $[^\circ]$         & 0.25 & 0.98 & 5.18 & 4.84 & 5.74 & 4.98  \\ \cline{2-8}
& Local RMS polar error $[^\circ]$  & 26.49 & 25.86 & 25.60 & 24.91 & 21.08 & 20.57 \\ \cline{2-8}
& Quadrant error {[}\%{]}           & 10.11 & 9.67 & 11.48 & 10.35 & 10.12 & 9.07 \\ \hline
\end{tabular}
\begin{tabular}{|l|l|c|c|c|c|c|c|}
  \multicolumn{8}{c}{\textbf{Second room} -   $RT60=1.88s$} \\
\hline
\multicolumn{2}{|r|}{HRTF dataset} & \multicolumn{2}{c|}{KEMAR MIT} & \multicolumn{2}{c|}{CIPIC SBJ3} & \multicolumn{2}{c|}{ARI-BTE CI1} \\ \hline
\multicolumn{2}{|r|}{Position} & A & B & A & B & A & B  \\ \hline \hline
\parbox[t]{2mm}{\multirow{5}{*}{\rotatebox[origin=c]{90}{Metrics}}} &
Lateral Bias $[^\circ]$             & -0.51 & 0.89 & -0.60 & 0.25 & -1.58 & -0.35 \\ \cline{2-8}
& Lateral RMS error $[^\circ]$      & 8.79 & 8.93 & 9.57 & 9.43 & 9.00 & 8.83 \\ \cline{2-8}
& Elevation bias $[^\circ]$         & 3.22 & 5.28 & 9.70 & 9.42 & 13.22 & 13.80 \\ \cline{2-8}
& Local RMS polar error $[^\circ]$  & 33.83 & 34.31 & 31.51 & 31.36 & 31.02 & 30.83 \\ \cline{2-8}
& Quadrant error {[}\%{]}           & 19.10 & 15.91 & 21.65 & 21.68 & 23.25 & 23.49\\ \hline
\end{tabular}
\end{center}
\label{tab:data}
\end{table*}
\begin{figure}[t]
    \vspace{-5mm}
    \centering
    \includegraphics[width=1\linewidth]{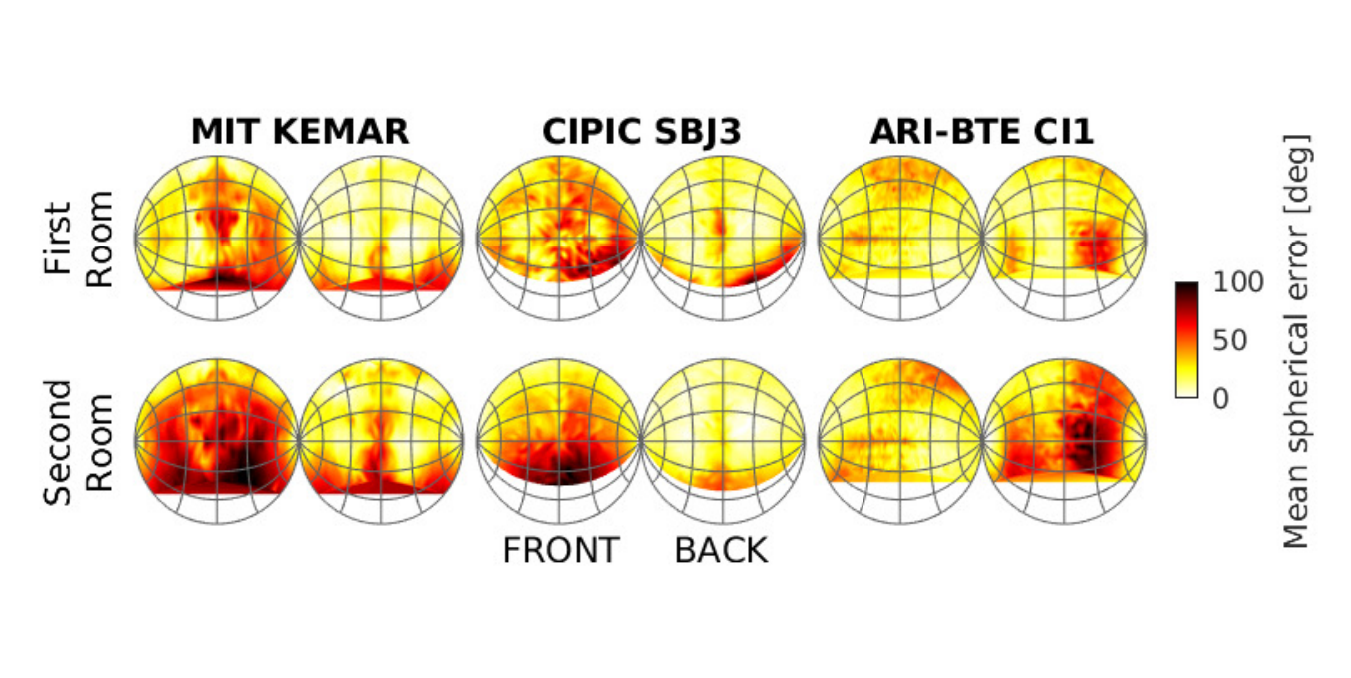}
    \caption{Mean average error over all accounted direction for the position A for both rooms. The plotted areas do not match since each HRTF dataset was acquired with different spatial grids.}
    \label{fig:plot_hrtf}
\end{figure}
Table \ref{tab:data} reports values of the above metrics over the different conditions, while Fig. \ref{fig:plot_hrtf} visualizes the simulations for position A and for all the evaluated HRTF datasets. The variability in the metrics values with respect to different positions and receiver types (i.e., HRTF sets) can be clearly appreciated. In particular, these results show that the model is sensitive to the spectral variations of the spatial cues when the receiver is located near a wall, as in position A. Finally, when increasing the reverberation time every simulated HRTF set underwent a performance degradation. This is due to the increase of multiple reflections that reduced the reliability of the localization cues. % aggiungere discussione dei risultati: il modello va male perchè completamente statico e media sul tempo, no head rotation and no precedence effect, contenuto in frequenza dei target molto diversi dal corrispettivo template, commentare se c'è qualche dipendenza della posizione ricevitore sorgente (la posizione centrale, C, dovrebbe essere la migliore in termini di precisione invece no) 
% Something related to the difference between first and second room missing....a simple sentence.

%=====================================
\section{Impact}
\label{sect:impact}
%\paper{
%\textbf{This is the main section of the article and the reviewers weight the description here appropriately}
%Indicate in what way new research questions can be pursued as a result of the software (if any).
%Indicate in what way, and to what extent, the pursuit of existing research questions is improved (if so).
%Indicate in what way the software has changed the daily practice of its users (if so).
%Indicate how widespread the use of the software is within and outside the intended user group.
%Indicate in what way the software is used in commercial settings and/or how it led to the creation of spin-off companies (if so).
%}
%Acoustic simulators has been demonstrated to be effective methodology to generate ad-hoc datasets to train and evaluate data-driven algorithms, i.e.~\cite{saurabh2017hearing}. Moreover, such 
Data-driven and machine-hearing systems are becoming a key component in audio signal processing research, but they require large amount of labelled data in order to be deployed~\cite{kataria_hearing_2017}. On the other hand, the process of recording such data from a real environment involves manual operations that are time consuming and error-prone. Datasets recorded in reverberant environments are particularly important to assess data-driven algorithms in real-world conditions, since anechoic sound can only be achieved in controlled environments~\cite{gannot2017consolidated}. While simulation tools cannot guarantee the same accuracy and realism of recorded data, they enable the user to assess the perfomance of a machine learning algorithm over a wider set of conditions %by adopting the \texttt{SofaMyRoom} for 
and with a rapid prototyping pipeline (consider Sect.~V.F of ~\cite{bianco_machine_2019}). Consequently, the adoption of simulation tools for producing artificial RIRs has been proposed in several research fields, e.g. speech enhancement and source separation~\cite{gannot2017consolidated}, or sound source localization~\cite{chakrabarty_broadband_2017}. A demonstrative examples on how to train several machine learning algorithms with synthetic spatial RIRs and then testing them with recorded dataset is reported in the work of Perotin et al.~\cite{perotin_regression_2019} while He et al.~\cite{he_adaptation_2019} extend this approach through domain adaptation.\todo{sistemare... probabilimente anche l'abstract va rivisto} Moreover, data augmentation techniques, especially targeted at the training of deep networks, are becoming popular even in the audio domain, and can help networks generalize their modeling capabilities~\cite{salamon2017deep}.
%non mi ricordo da dove arriva sta roba \todo{offline reinforcement learning \url{https://proceedings.neurips.cc/paper/2020/file/f7efa4f864ae9b88d43527f4b14f750f-Paper.pdf}}

Acoustic simulation tools can be profitably used for all these purposes, and can change drastically the development workflow of a machine hearing system. Particularly, the developer can leverage the flexibility of such tools to iteratively generate and evaluate data (i.e. with increasing size, complexity, number of conditions) along the development process. 
%More in detail, it is possible to start the training procedure on a basic set of conditions to test the correctness of the algorithm under test and, in a second moment, increase the complexity of the test conditions to further benchmark the learning framework. 
A similar workflow is in principle possible also when adopting a recorded dataset, but is typically unfeasible in practice because of limitations in the available data or heterogeneity of formats.
%its specific conditions, it can be necessary to integrate a second dataset to evaluate the algorithm on conditions that the first one did not foresee. Furthermore, find the proper dataset is itself a difficult task since: the conditions required can be highly specific shrinking the usable datasets, the dataset could require a usage fee, the structure could require third-part software or a specific code snippet to access the data ecc. With a simulation tool such as \texttt{SofaMyRoom}, it is possible to simulate uncommon conditions while maintaining the same data structure thus enabling a seamless integration into the learning system development. 

\texttt{SofaMyRoom} aims at filling a gap in the availability of open-source, fully configurable, and (relatively) user-friendly room acoustics simulators. 
Although commercial solutions exist providing accurate simulations for complex acoustic geometries, they have the disadvantage of being closed, and entail non-negligible financial costs. On the other hand, none of the existing academic projects integrates into a single tool all the functionalities and features provided by \texttt{SofaMyRoom}. 

By virtue of the functionalities discussed in the previous sections, we expect it to be a useful tool to generate labelled data for several research problems that consider the acoustic and psychoacoustic effects of echoic environments and binaural listening. 
Althought synthetic BRIRs generated from simple geometries lead to reduced perceptual plausibility, tools like \texttt{SofaMyRoom} have been easily integrated in the stimulus generation pipeline in order to address several research challenges i.e.: (i) subjective experiments on speech understanding and perceived reverberation~\cite{ellis_dissociation_2019}, (ii) supporting hearing aids simulations~\cite{mueller_localization_2012} or (iii) testing novel on-board algorithms for cochlear implants~\cite{lopez-poveda_lateralization_2019}. It is worthwhile to notice that adopting our publicly available tool and by making simulation parameters freely accessible will grant reproducible research by allowing the re-synthesis of the stimuli.
%, including: sound localization, distance estimation, automatic speech recognition, speech separation and dereverberation, estimation of acoustic parameters.

%As a conclusion and to support the idea that this toolbox can be an organic element for audio research, the use of \texttt{SofaMyRoom} can also find a key role in other research fields:
Finally, in addition to psycho-acoustics and machine hearing, \texttt{SofaMyRoom} can be employed in further audio-related research fields. As an example, in auditory modelling research~\cite{lavandier_prediction_2010} it may be used to provide specific acoustic conditions to understand the limitations of the model or, in robot audition research~\cite{deleforge2015acoustic}, it can provide a simulated environment to benchmark the robot performances.

%While the toolbox at this stage of development is mainly intended to return only monophonic or stereophonic impulse responses, is worthwhile to notice that the SOFA format can potentially integrate multi-channel microphone arrays expanding the possible range of research topics. 

One current limitation of \texttt{SofaMyRoom} is that it only considers ``shoebox'' shaped rooms. This prevents to simulate the complexity of a real acoustic space, which however is not the goal of this project. Instead it provides an efficient solution for the use cases outlined in the manuscript: as an example, shoe-box room shapes have been used to train deep networks in some recent works on sound source localization~\cite{perotin_regression_2019,carlo_mirage_2019}, that hint at the ability of the networks to generalize to real-world data despite the use of simplified geometries in the training stage.

One potential pitfall of an open-source projects is maintainability and portability over different platforms. As far as dependency on external libraries is concerned, we have relied on code that is well maintained and actively developed. In particular, the SOFA format is increasingly adopted within the research community and the industry.\footnote{As an example, see https://www.genelec.com/aural-id} This is why our effort aimed at releasing a reliable simulation tool that can be easily maintained and extended thanks to its building system that provides a multi platform compilation~\cite{wilson_best_2014}. 

%, we believe that this toolbox will be a valuable tool for the audio signal processing community.

%=====================================
\section{Conclusions}
\label{sect:conclusions}
%\paper{Set out the conclusion of this original software publication.}
This paper has reported on the development of \texttt{SofaMyRoom}, a room acoustics simulator which can render binaural room impulse responses given different parameters for the room, receiver and sound source. Details on the software architecture, functionalities, and set-up, were provided. The employed numerical methods were also briefly discussed. By means of a proof of concept example, we showed how \texttt{SofaMyRoom} can be used into a machine hearing workflow. We believe that the proposed framework can be useful for several human and machine audition's challenges such as: cocktail party simulations, localization in highly reverberant scenarios and speech in noise. Improvements in the near future will be mainly aimed at providing support to multi-channel microphone array receivers, also considered in the SOFA format, and at developing a Python wrapper that is expected to boost wider adoption of the simulator.

\section{Conflict of Interest}
We wish to confirm that there are no known conflicts of interest associated with this publication and there has been no significant financial support for this work that could have influenced its outcome.

\section*{Acknowledgements}

We thank Clément Gaultier, Antonie Deleforge and Diego Di Carlo for providing the code of the VAST project. 

Icons in Figure \ref{fig:sofamyroomschema} were downloaded and adapted from www.flaticon.com (Authors: wanicon, 
Kiranshastry, freepik, surang)
    

%% References:
\bibliographystyle{elsarticle-num} 
\bibliography{main.bib}

\begin{thebibliography}{10}
\expandafter\ifx\csname url\endcsname\relax
  \def\url#1{\texttt{#1}}\fi
\expandafter\ifx\csname urlprefix\endcsname\relax\def\urlprefix{URL }\fi
\expandafter\ifx\csname href\endcsname\relax
  \def\href#1#2{#2} \def\path#1{#1}\fi

\bibitem{kuttruff2016room}
H.~Kuttruff, Room acoustics, CRC Press, 2016.

\bibitem{blauert2013technology}
J.~Blauert, The technology of binaural listening, Springer, 2013.

\bibitem{allen_image_1979}
J.~B. Allen, D.~A. Berkley, Image method for efficiently simulating small-room
  acoustics, The Journal of the Acoustical Society of America 65~(4) (1979)
  943--950.

\bibitem{lyon2017human}
R.~F. Lyon, Human and machine hearing, Cambridge University Press, 2017.

\bibitem{haykin2005cocktail}
S.~Haykin, Z.~Chen, The cocktail party problem, Neural computation 17~(9)
  (2005) 1875--1902.

\bibitem{gannot2017consolidated}
S.~Gannot, E.~Vincent, S.~Markovich-Golan, A.~Ozerov, A consolidated
  perspective on multimicrophone speech enhancement and source separation,
  IEEE/ACM Transactions on Audio, Speech, and Language Processing 25~(4) (2017)
  692--730.

\bibitem{brown1994computational}
G.~J. Brown, M.~Cooke, Computational auditory scene analysis, Computer speech
  and language 8~(4) (1994) 297--336.

\bibitem{gardner_hrtf_1995}
W.~G. Gardner, K.~D. Martin,
  \href{http://asa.scitation.org/doi/abs/10.1121/1.412407}{{HRTF} measurements
  of a {KEMAR}}, The Journal of the Acoustical Society of America 97~(6) (1995)
  3907--3908.
\newblock \href {https://doi.org/10.1121/1.412407}
  {\path{doi:10.1121/1.412407}}.
\newline\urlprefix\url{http://asa.scitation.org/doi/abs/10.1121/1.412407}

\bibitem{schimmel_fast_2009}
S.~M. Schimmel, M.~F. Muller, N.~Dillier, A fast and accurate “shoebox”
  room acoustics simulator, in: 2009 {IEEE} International Conference on
  Acoustics, Speech and Signal Processing, 2009, pp. 241--244.
\newblock \href {https://doi.org/10.1109/ICASSP.2009.4959565}
  {\path{doi:10.1109/ICASSP.2009.4959565}}.

\bibitem{heinz_binaural_1993}
R.~Heinz,
  \href{https://linkinghub.elsevier.com/retrieve/pii/0003682X9390048B}{Binaural
  room simulation based on an image source model with addition of statistical
  methods to include the diffuse sound scattering of walls and to predict the
  reverberant tail}, Applied Acoustics 38~(2) (1993) 145--159.
\newblock \href {https://doi.org/10.1016/0003-682X(93)90048-B}
  {\path{doi:10.1016/0003-682X(93)90048-B}}.
\newline\urlprefix\url{https://linkinghub.elsevier.com/retrieve/pii/0003682X9390048B}

\bibitem{majdak_spatially_2013}
P.~Majdak, Y.~Iwaya, T.~Carpentier, R.~Nicol, M.~Parmentier, A.~Roginska,
  Y.~Suzuki, K.~Watanabe, H.~Wierstorf, H.~Ziegelwanger, M.~Noisternig,
  \href{http://www.aes.org/e-lib/browse.cfm?elib=16781}{Spatially oriented
  format for acoustics: A data exchange format representing head-related
  transfer functions}, in: Audio Engineering Society Convention 134, Audio
  Engineering Society, 2013, pp. 1--11.
\newline\urlprefix\url{http://www.aes.org/e-lib/browse.cfm?elib=16781}

\bibitem{geronazzo_standardized_2013}
M.~Geronazzo, F.~Granza, S.~Spagnol, F.~Avanzini,
  \href{http://www.aes.org/e-lib/browse.cfm?elib=16802}{A standardized
  repository of head-related and headphone impulse response data}, in: Proc.
  134th {Conv}. {Audio} {Eng}. {Society}, Rome, Italy, 2013, pp. 1--7.
\newline\urlprefix\url{http://www.aes.org/e-lib/browse.cfm?elib=16802}

\bibitem{gaultier2017vast}
C.~Gaultier, S.~Kataria, A.~Deleforge,
  \href{https://hal.archives-ouvertes.fr/hal-01416508}{{VAST : The Virtual
  Acoustic Space Traveler Dataset}}, in: {International Conference on Latent
  Variable Analysis and Signal Separation (LVA/ICA)}, Grenoble, France, 2017,
  pp. 1--10.
\newline\urlprefix\url{https://hal.archives-ouvertes.fr/hal-01416508}

\bibitem{dong_data_2018}
X.~L. Dong, T.~Rekatsinas,
  \href{http://dl.acm.org/citation.cfm?doid=3183713.3197387}{Data {Integration}
  and {Machine} {Learning}: {A} {Natural} {Synergy}}, in: Proceedings of the
  2018 {International} {Conference} on {Management} of {Data} - {SIGMOD} '18,
  ACM Press, Houston, TX, USA, 2018, pp. 1645--1650.
\newblock \href {https://doi.org/10.1145/3183713.3197387}
  {\path{doi:10.1145/3183713.3197387}}.
\newline\urlprefix\url{http://dl.acm.org/citation.cfm?doid=3183713.3197387}

\bibitem{hoene_mysofa_2017}
C.~Hoene, I.~C. Patino~Mejia, A.~Cacerovschi, Mysofa—design your personal
  hrtf, in: Audio Engineering Society Convention 142, Audio Engineering
  Society, 2017, pp. 1--6.

\bibitem{kataria_hearing_2017}
S.~Kataria, C.~Gaultier, A.~Deleforge,
  \href{https://hal.inria.fr/hal-01372435}{{Hearing in a shoe-box : binaural
  source position and wall absorption estimation using virtually supervised
  learning }}, in: {2017 IEEE International Conference on Acoustics, Speech and
  Signal Processing (ICASSP)}, New-Orleans, United States, 2017, pp. 226--230.
\newline\urlprefix\url{https://hal.inria.fr/hal-01372435}

\bibitem{reijniers_ideal-observer_2014}
J.~Reijniers, D.~Vanderelst, C.~Jin, S.~Carlile, H.~Peremans,
  \href{https://doi.org/10.1007/s00422-014-0588-4}{An ideal-observer model of
  human sound localization}, Biological Cybernetics 108~(2) (2014) 169--181.
\newblock \href {https://doi.org/10.1007/s00422-014-0588-4}
  {\path{doi:10.1007/s00422-014-0588-4}}.
\newline\urlprefix\url{https://doi.org/10.1007/s00422-014-0588-4}

\bibitem{barumerli_model_2020}
R.~Barumerli, P.~Majdak, R.~Baumgartner, J.~Reijniers, M.~Geronazzo,
  F.~Avanzini, Predicting directional sound-localization of human listeners in
  both horizontal and vertical dimensions, in: Audio Engineering Society
  Convention 148, Audio Engineering Society, 2020, p.~8.

\bibitem{sondergaard_auditory_2013}
P.~L. Søndergaard, P.~Majdak,
  \href{https://link.springer.com/chapter/10.1007/978-3-642-37762-4_2}{The
  auditory modeling toolbox}, in: The Technology of Binaural Listening,
  Springer, Berlin, Heidelberg, 2013, pp. 33--56.
\newblock \href {https://doi.org/10.1007/978-3-642-37762-4_2}
  {\path{doi:10.1007/978-3-642-37762-4_2}}.
\newline\urlprefix\url{https://link.springer.com/chapter/10.1007/978-3-642-37762-4_2}

\bibitem{barumerli2019auditory}
R.~Barumerli, A.~Almenari, M.~Geronazzo, G.~M. Di~Nunzio, F.~Avanzini, Auditory
  models comparison for horizontal localization of concurrent speakers in
  adverse acoustic scenarios, in: 23rd International Congress on Acoustics,
  2019, pp. 7686--7693.

\bibitem{algazi_cipic_2001}
V.~R. Algazi, R.~O. Duda, D.~M. Thompson, C.~Avendano, The {CIPIC} {HRTF}
  database, in: Proceedings of the 2001 {IEEE} Workshop on the Applications of
  Signal Processing to Audio and Acoustics (Cat. No.01TH8575), 2001, pp.
  99--102.
\newblock \href {https://doi.org/10.1109/ASPAA.2001.969552}
  {\path{doi:10.1109/ASPAA.2001.969552}}.

\bibitem{majdak_two-dimensional_2011}
P.~Majdak, M.~J. Goupell, B.~Laback,
  \href{http://journals.lww.com/00003446-201103000-00005}{Two-{Dimensional}
  {Localization} of {Virtual} {Sound} {Sources} in {Cochlear}-{Implant}
  {Listeners}:}, Ear and Hearing 32~(2) (2011) 198--208.
\newblock \href {https://doi.org/10.1097/AUD.0b013e3181f4dfe9}
  {\path{doi:10.1097/AUD.0b013e3181f4dfe9}}.
\newline\urlprefix\url{http://journals.lww.com/00003446-201103000-00005}

\bibitem{middlebrooks_virtual_1999}
J.~C. Middlebrooks,
  \href{http://asa.scitation.org/doi/abs/10.1121/1.427147}{Virtual localization
  improved by scaling nonindividualized external-ear transfer functions in
  frequency}, The Journal of the Acoustical Society of America 106~(3) (1999)
  1493--1510.
\newblock \href {https://doi.org/10.1121/1.427147}
  {\path{doi:10.1121/1.427147}}.
\newline\urlprefix\url{http://asa.scitation.org/doi/abs/10.1121/1.427147}

\bibitem{bianco_machine_2019}
M.~J. Bianco, P.~Gerstoft, J.~Traer, E.~Ozanich, M.~A. Roch, S.~Gannot, C.-A.
  Deledalle, \href{http://asa.scitation.org/doi/10.1121/1.5133944}{Machine
  learning in acoustics: {Theory} and applications}, The Journal of the
  Acoustical Society of America 146~(5) (2019) 3590--3628.
\newblock \href {https://doi.org/10.1121/1.5133944}
  {\path{doi:10.1121/1.5133944}}.
\newline\urlprefix\url{http://asa.scitation.org/doi/10.1121/1.5133944}

\bibitem{chakrabarty_broadband_2017}
S.~Chakrabarty, E.~A.~P. Habets,
  \href{http://ieeexplore.ieee.org/document/8170010/}{Broadband doa estimation
  using convolutional neural networks trained with noise signals}, in: 2017
  {IEEE} {Workshop} on {Applications} of {Signal} {Processing} to {Audio} and
  {Acoustics} ({WASPAA}), IEEE, New Paltz, NY, 2017, pp. 136--140.
\newblock \href {https://doi.org/10.1109/WASPAA.2017.8170010}
  {\path{doi:10.1109/WASPAA.2017.8170010}}.
\newline\urlprefix\url{http://ieeexplore.ieee.org/document/8170010/}

\bibitem{perotin_regression_2019}
L.~Perotin, A.~Defossez, E.~Vincent, R.~Serizel, A.~Guerin,
  \href{https://ieeexplore.ieee.org/document/8937277/}{Regression {Versus}
  {Classification} for {Neural} {Network} {Based} {Audio} {Source}
  {Localization}}, in: 2019 {IEEE} {Workshop} on {Applications} of {Signal}
  {Processing} to {Audio} and {Acoustics} ({WASPAA}), IEEE, New Paltz, NY, USA,
  2019, pp. 343--347.
\newblock \href {https://doi.org/10.1109/WASPAA.2019.8937277}
  {\path{doi:10.1109/WASPAA.2019.8937277}}.
\newline\urlprefix\url{https://ieeexplore.ieee.org/document/8937277/}

\bibitem{he_adaptation_2019}
W.~He, P.~Motlicek, J.-M. Odobez,
  \href{https://ieeexplore.ieee.org/document/8682655/}{Adaptation of {Multiple}
  {Sound} {Source} {Localization} {Neural} {Networks} with {Weak} {Supervision}
  and {Domain}-adversarial {Training}}, in: {ICASSP} 2019 - 2019 {IEEE}
  {International} {Conference} on {Acoustics}, {Speech} and {Signal}
  {Processing} ({ICASSP}), IEEE, Brighton, United Kingdom, 2019, pp. 770--774.
\newblock \href {https://doi.org/10.1109/ICASSP.2019.8682655}
  {\path{doi:10.1109/ICASSP.2019.8682655}}.
\newline\urlprefix\url{https://ieeexplore.ieee.org/document/8682655/}

\bibitem{salamon2017deep}
J.~Salamon, J.~P. Bello, Deep convolutional neural networks and data
  augmentation for environmental sound classification, IEEE Sig. Process.
  Letters 24~(3) (2017) 279--283.

\bibitem{ellis_dissociation_2019}
G.~M. Ellis, P.~Zahorik,
  \href{https://linkinghub.elsevier.com/retrieve/pii/S0378595518305112}{A
  dissociation between speech understanding and perceived reverberation},
  Hearing Research 379 (2019) 52--58.
\newblock \href {https://doi.org/10.1016/j.heares.2019.04.015}
  {\path{doi:10.1016/j.heares.2019.04.015}}.
\newline\urlprefix\url{https://linkinghub.elsevier.com/retrieve/pii/S0378595518305112}

\bibitem{mueller_localization_2012}
M.~F. Mueller, A.~Kegel, S.~M. Schimmel, N.~Dillier, M.~Hofbauer,
  \href{http://asa.scitation.org/doi/10.1121/1.4705292}{Localization of virtual
  sound sources with bilateral hearing aids in realistic acoustical scenes},
  The Journal of the Acoustical Society of America 131~(6) (2012) 4732--4742.
\newblock \href {https://doi.org/10.1121/1.4705292}
  {\path{doi:10.1121/1.4705292}}.
\newline\urlprefix\url{http://asa.scitation.org/doi/10.1121/1.4705292}

\bibitem{lopez-poveda_lateralization_2019}
E.~A. Lopez-Poveda, A.~Eustaquio-Martín, M.~J. Fumero, J.~S. Stohl,
  R.~Schatzer, P.~Nopp, R.~D. Wolford, J.~M. Gorospe, R.~Polo, A.~G. Revilla,
  B.~S. Wilson,
  \href{https://linkinghub.elsevier.com/retrieve/pii/S0378595518305355}{Lateralization
  of virtual sound sources with a binaural cochlear-implant sound coding
  strategy inspired by the medial olivocochlear reflex}, Hearing Research 379
  (2019) 103--116.
\newblock \href {https://doi.org/10.1016/j.heares.2019.05.004}
  {\path{doi:10.1016/j.heares.2019.05.004}}.
\newline\urlprefix\url{https://linkinghub.elsevier.com/retrieve/pii/S0378595518305355}

\bibitem{lavandier_prediction_2010}
M.~Lavandier, J.~F. Culling,
  \href{http://asa.scitation.org/doi/10.1121/1.3268612}{Prediction of binaural
  speech intelligibility against noise in rooms}, The Journal of the Acoustical
  Society of America 127~(1) (2010) 387--399.
\newblock \href {https://doi.org/10.1121/1.3268612}
  {\path{doi:10.1121/1.3268612}}.
\newline\urlprefix\url{http://asa.scitation.org/doi/10.1121/1.3268612}

\bibitem{deleforge2015acoustic}
A.~Deleforge, F.~Forbes, R.~Horaud, Acoustic space learning for sound-source
  separation and localization on binaural manifolds, International journal of
  neural systems 25~(01) (2015) 1440003.

\bibitem{carlo_mirage_2019}
D.~Di~Carlo, A.~Deleforge, N.~Bertin,
  \href{https://ieeexplore.ieee.org/document/8683534/}{Mirage: {2D} {Source}
  {Localization} {Using} {Microphone} {Pair} {Augmentation} with {Echoes}}, in:
  {ICASSP} 2019 - 2019 {IEEE} {International} {Conference} on {Acoustics},
  {Speech} and {Signal} {Processing} ({ICASSP}), IEEE, Brighton, United
  Kingdom, 2019, pp. 775--779.
\newblock \href {https://doi.org/10.1109/ICASSP.2019.8683534}
  {\path{doi:10.1109/ICASSP.2019.8683534}}.
\newline\urlprefix\url{https://ieeexplore.ieee.org/document/8683534/}

\bibitem{wilson_best_2014}
G.~Wilson, D.~A. Aruliah, C.~T. Brown, N.~P.~C. Hong, M.~Davis, R.~T. Guy,
  S.~H.~D. Haddock, K.~D. Huff, I.~M. Mitchell, M.~D. Plumbley, B.~Waugh, E.~P.
  White, P.~Wilson,
  \href{https://journals.plos.org/plosbiology/article?id=10.1371/journal.pbio.1001745}{Best
  {Practices} for {Scientific} {Computing}}, PLOS Biology 12~(1) (2014)
  e1001745, publisher: Public Library of Science.
\newblock \href {https://doi.org/10.1371/journal.pbio.1001745}
  {\path{doi:10.1371/journal.pbio.1001745}}.
\newline\urlprefix\url{https://journals.plos.org/plosbiology/article?id=10.1371/journal.pbio.1001745}

\end{thebibliography}


%\begin{thebibliography}{00}


%\bibitem{}

%\end{thebibliography}
\newpage
\section*{Current executable software version}
\begin{table}[!h]
\begin{tabular}{|l|p{6.5cm}|p{6.5cm}|}
\hline
\textbf{Nr.} & \textbf{(Executable) software metadata description} & \textbf{Please fill in this column} \\
\hline
S1 & Current software version & 1.0 \\
\hline
S2 & Permanent link to executables of this version  & \url{https://github.com/spatialaudiotools/sofamyroom/archive/master.zip}\\
\hline
S3 & Legal Software License & EUPL v1.2 \\
\hline
S4 & Computing platforms/Operating Systems & Apple macOS, Ubuntu 20.04, Microsoft Windows 10\\
\hline
S5 & Installation requirements \& dependencies & Only for Linux based systems \\
\hline
S6 & If available, link to user manual - if formally published include a reference to the publication in the reference list & \url{https://spatialaudiotools.github.io/sofamyroom/} \\
\hline
S7 & Support email for questions & \\
\hline
\end{tabular}
\caption{Software metadata}
 
\end{table}

\end{document}